#\documentstyle[psfig,epsf]{/disk/3/roud/home/Texstuff/aa}
\documentstyle[psfig,epsf]{aa}

\newcommand{\bb}{\bibitem[]{bla}}
\newcommand{\zm}{ \relax \ifmmode {\rm M_{\odot}} \else {M$_{\odot}$}\fi}

\newcommand{\degree}{$^{\rm o}$}

\newcommand{\ea}{{et al.}}

\def\Mv{\mbox{$M_{_V}$}}

\def\lesssim{\mathrel{\hbox{\rlap{\hbox{\lower4pt\hbox{$\sim$}}}\hbox{$<$}}}}
\def\gtrsim{\mathrel{\hbox{\rlap{\hbox{\lower4pt\hbox{$\sim$}}}\hbox{$>$}}}}

\def\ion#1#2{#1$\;${\small\rm\@Roman{#2}}\relax}
 
\newcommand{\spi}{$\sigma/\pi$}
\begin{document}

   \thesaurus{ 
              (08.04.1;  
               08.06.3;  
               08.12.1)  }

\title{
The absolute magnitude of K0V stars from {\sc hipparcos} parallaxes
}

\author{
Ren\'e D. Oudmaijer \inst{1}, Martin A.T. Groenewegen \inst{2} 
\& Hans Schrijver \inst{3} 
}

\offprints{Ren\'e Oudmaijer}

\institute{Astrophysics Group, Imperial College of Science, 
Technology and Medicine, Prince Consort Road, London SW7 2BZ, U.K. 
\and
Max-Planck-Institut f\"ur Astrophysik, 
Karl-Schwarzschild-Stra{\ss}e 1, D-85740 Garching, Germany 
\and
SRON, Sorbonnelaan 2, NL-3584 CA Utrecht, The Netherlands \\
}

\date{Accepted 1999.
      Received 1999;
      in original form 1999}

\authorrunning{Oudmaijer, Groenewegen \& Schrijver}
\titlerunning{Absolute magnitude of K0V stars}

\maketitle

\begin{abstract}

We investigate the properties of K0V stars with Hipparcos parallaxes
and spectral types taken from the Michigan Spectral Survey.  The
sample of 200 objects allows the empirical investigation of the
magnitude selection (Malmquist) bias, which appears clearly present.
By selecting those objects that are not affected by bias, we find a
mean absolute magnitude of \Mv~=~5.7, a downward revision from 5.9
mag.  listed in Schmidt-Kaler (1982).  Some objects have absolute
magnitudes far brighter than \Mv~=~5.7, and it is suggested that these
objects ($\approx$~20\% of the total sample) are K0IV stars which may
have been mis-classified as a K0V star.  The presence of the Malmquist
bias in even this high quality sample suggests that no sample can be
expected to be bias-free.

 
\end{abstract}

\section{Introduction}

The Hipparcos trigonometric parallax measurements of more than 100 000
stars (ESA, 1997) provide an excellent basis to determine the
fundamental parameters of stars.  Yet, some, not always trivial,
problems arise which make the conversion from the measured parallax to
intrinsic absolute magnitude of an object not
straightforward.  For example, the Lutz-Kelker effect (Lutz \& Kelker,
1973), results in too faint magnitudes for large relative errors \spi,
while the Malmquist bias results in too bright mean absolute
magnitudes, because at the observed magnitude limit, brighter objects
will be included in a sample, while fainter objects will not.
Additional complications are listed in Brown \ea\ (1997).

In a previous paper (Oudmaijer, Groenewegen \& Schrijver 1998 -
hereafter OGS98) we have shown empirically that the Lutz-Kelker bias
is present in trigonometric parallaxes.  This was done by comparing
the best Hipparcos parallaxes (\spi~$<$~5\% -- defining a
`true' parallax sample) with lower quality ground-based parallaxes of
a large sample of stars.  The data showed that, for increasing \spi ,
the derived absolute magnitude of an object indeed becomes too faint
in a manner consistent with the Lutz-Kelker predictions (see also Koen
1992), but for even larger \spi , the derived magnitudes became too
bright  by up to 2 magnitudes.  The sample was evidently not
hampered by only one type of bias, but by at least two. The first being
the Lutz-Kelker bias, the second  was called the `completeness
effect', which we now identify  as the magnitude selection Malmquist
bias.

To investigate this further, we tackle the problem in a similar way by
analyzing a sample of stars for which it may be hoped that all
have  the same intrinsic magnitude.  To this end, we have
investigated a sample of stars with well-defined spectral
types, the K0V stars.

\section{Sample selection}

To determine the absolute magnitudes of stars with the same spectral
type, a coherent and homogeneous database of spectral types is needed. 
The Michigan Spectral Survey Volumes 1..4 (MSS, Houk \& Cowley 1975;
Houk 1978; Houk 1982 and Houk \& Smith-Moore 1988, providing
spectral types of the HD catalogue in Declination from --90\degree \ to
--12\degree) is such a database.  We chose to investigate K0V stars, as
these objects are relatively close by and will not suffer much from
interstellar extinction, while the number of objects is relatively
large.  The selection criteria from the Hipparcos  Catalogue (ESA,
1997) were: 

\smallskip
\noindent
(i) Spectral type = `K0V ' (Field H76, the sources for the spectral 
types are listed in Field H77)\\
\noindent
(ii) Goodness-of-fit $<$ 3   (Field H29) \\
\noindent
(iii) Number of rejected data $<$ 10\% (Field H30) 

\smallskip

563 objects in the Hipparcos catalogue have `K0V ' listed in their
spectral type entry, but the majority of the stars has the
spectral type taken from other sources than the MSS or are listed as
`G8V/K0V'. These objects were rejected, leaving a sample of 201
objects. Only one of these has a negative parallax (HD 219882), and
was also rejected for further analysis, one object (HD 170132) has no
{\it (B--V)} listed in the Hipparcos catalogue, its value was taken from
the {\sc simbad} database.

The selection thus yielded 200 objects.  The average error on the
parallax and its scatter, are 1.4~$\pm$~0.6 mas, and the bulk of the
sample has parallaxes larger than 10 mas, probing the nearest 100 pc.
The quality of the parallaxes is extremely high; the average \spi \ is
11\%, indicating a 9$\sigma$ detection on average.

\section{Properties of the sample}

\begin{figure}
\mbox{\epsfxsize=0.45\textwidth\epsfbox[32 171 512 647]
{
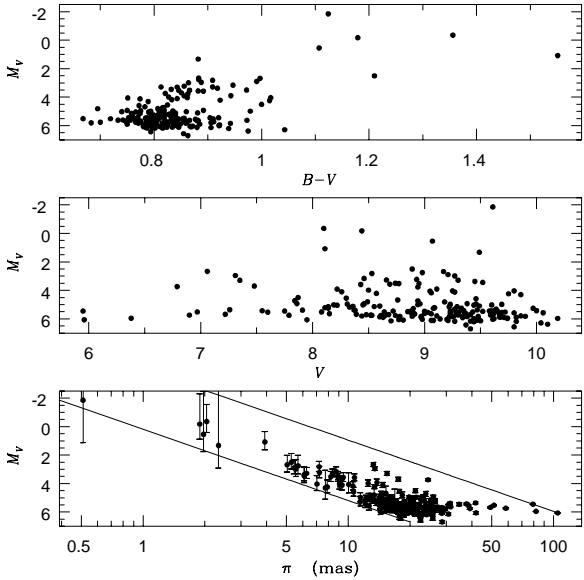}}
\caption{Derived \Mv \ as function of several parameters.
Errorbars on the absolute magnitudes are for
convenience only shown in the lower panel, and are often smaller than
the plotsymbols.  The solid lines are drawn
according to Eq.~1 in OGS98, and explained in the text.
\label{kovf1}
}
\end{figure}

The absolute magnitude, derived from the parallax and the {\it V} magnitude,
neglecting interstellar extinction, is plotted in Fig.~\ref{kovf1}.
The upper panel shows \Mv \ against {\it (B--V)}.  The unweighted mean
\Mv~=~5.06~$\pm$~1.26 (the r.m.s.  deviation around the mean) is
almost 1 magnitude brighter compared to what is expected for K0V stars
(\Mv~=~5.9, Schmidt-Kaler 1982, hereafter SK82).  Some objects are
even 6 - 8 magnitudes brighter than a normal K0V star.  A trend in
({\it B--V}) may be present, as the redder stars correspond to the
intrinsically brightest objects.  The relation between \Mv \ and {\it
V} (middle panel) shows a large scatter which seems to increase for
fainter objects.  There is a strong correlation between the derived
\Mv \ and the measured parallax (lower panel).  For small parallaxes,
the intrinsic magnitude becomes brighter and, interestingly, for the
smallest parallaxes, no objects have intrinsic magnitudes that are
even close to \Mv~=~5.9.  A strong limit to the derived \Mv \
as function of parallax is present, which is due to the `completeness
effect' mentioned in OGS98.

As discussed in OGS98, the difference between the derived absolute
magnitude of an object from its parallax and the limiting observed {\it
V} magnitude of a sample define a `forbidden' region.  The reason is
that stars that would have been present in the fainter regions (the
lower left hand corner of the lower panel) are simply too faint to be
included in the sample.  An upper bound is also present, reflecting the
fact that fewer bright objects are present than faint objects. 
The solid lines in the figure indicate the regions where no data are
expected, and are drawn according to Eq.~1 in OGS98, with  limiting
magnitudes corresponding to the faintest  {\it V} magnitude in
this sample, {\it V}~=~10.19, respectively the brightest, 
{\it V}~=~5.95. 

\begin{figure}
\mbox{\epsfxsize=0.45\textwidth\epsfbox[38 171 516 425]
{
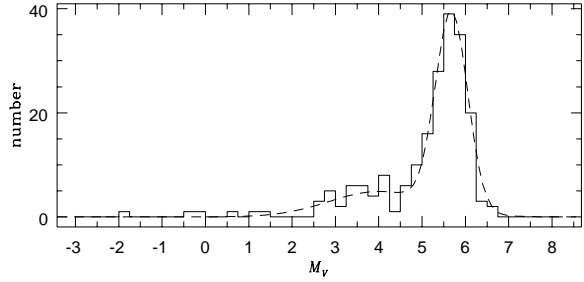}}
\caption{ The distribution of \Mv , binned to 0.25
mag.  bins. The dotted line represents a two component Gaussian fit to
the data.
\label{kovf2}
}
\end{figure}

We now identify this `completeness effect' with the magnitude
selection Malmquist bias, which is working on exactly the same
principle, and effectively forbids the use of entire samples to derive
their mean absolute magnitude, without correcting for it and/or
investigating when the bias starts to dominate.  It has been advocated
to first plot luminosities as a function of distance, a parameter
directly related to the distance such as the red-shift in the case of
galaxies (e.g.  Sandage 1994) or, in this case parallax, to assess the
presence of selection biases.  Such diagnostic plots, sometimes
called Spaenhauer diagrams after Spaenhauer (1978),
also serve to identify Lutz-Kelker type biases (see OGS98).
Sandage (1994) showed that his sample of galaxies suffers from the
Malmquist bias (with $\Delta$M~=~1.386 $\times$ $\sigma^2$, for a
uniform space distribution, with $\sigma$ the assumed intrinsic
scatter of the absolute magnitude distribution, see e.g.  Hanson,
1979) when he compared the average with a sub-sample, easily
identifiable in the diagrams, which is not affected.

Let us now derive the Malmquist correction for our sample,
Fig.~\ref{kovf1} shows that those objects with $\pi$~$>$~20 mas are
not affected -- the unweighted mean of these objects returns a value
of \Mv~=~5.69 with an (intrinsic) scatter of 0.4 mag.  The entire
sample returns an average of 5.06~$\pm$~1.26 mag. The expected
Malmquist correction is 0.22 mag for $\sigma$~=~0.4, so the difference
between the derived \Mv \ for the entire sample and that of the
unaffected sub-sample is much larger than what the Malmquist bias
predicts. This is at first sight hard to understand, but may be
related to the question why we would find K0V stars which are up to
6-8 magnitudes brighter than expected.  Apart from the rather unlikely
possibilities that parallax errors would result in such
deviant values (these objects have very high signal-to-noise
detections) or that the class of K0V stars can have such a large range
of intrinsic magnitudes, it may be more likely that the discrepancy is
due to spectral misclassification.

Some information may be gained from Fig.~\ref{kovf2}, where the
distribution of the derived \Mv \ values is shown.  The distribution
peaks close to 5.7, but is not symmetric around the mean; a secondary
maximum appears close to \Mv~=~3.5.  The presence of a secondary peak
strongly suggests that an additional population of stars is present.
These could be objects with a different spectral type as the peak
roughly agrees with the magnitudes of K0IV stars (\Mv~=~3.1, SK82),
while K0III giants may also be present  (\Mv~=~0.7,
SK82).  It is hard to make a good distinction between the different
groups, judging the gaps between the \Mv~$\approx$~5.7 and
\Mv~$\approx$~3.5 objects in Figs.~\ref{kovf1} and \ref{kovf2}, the
separation seems to be present for \Mv~=~4.5.

\begin{figure}
\mbox{\epsfxsize=0.45\textwidth\epsfbox[32 171 512 425]
{
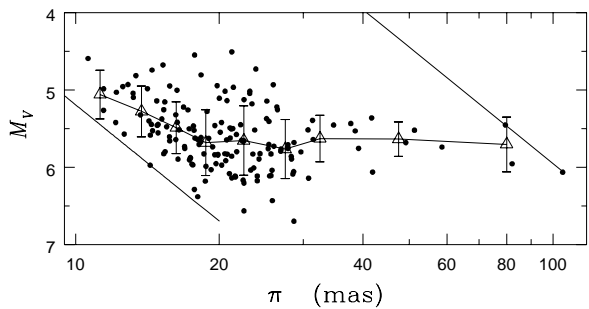}}
\caption{ The resulting K0V sample -- 159 objects with \Mv~$>$~4.5. The
solid lines are as in the previous figure. The triangles indicate
the mean and its scatter in the parallax bins 10-12.5; 12.5-15;
15-17.5; 17.5-20; 20-25; 25-30; 30-35; 35-60 and $>$~60 mas.
\label{kovf3}
}
\end{figure}

159 of the 200 objects are present in the `faint' sample with
\Mv~$>$~4.5, the remaining 41 stars have brighter intrinsic magnitudes. 
The average {\it (B--V)} of the faint (i.e.  K0V) sample is
0.82~$\pm$~0.05, consistent with the intrinsic colours for the group
(0.81, SK82), also suggesting that our neglect of interstellar reddening
is warranted.  The remaining objects have a redder average {\it (B--V)}
of 0.94 but with a large scatter of 0.16 mag.  If we reject the 6
brightest objects in this sample, the scatter is reduced and the average
{\it (B--V)}~=~0.89~$\pm$~0.08, with an average \Mv \ of 3.5~$\pm$~0.5,
consistent with a K0IV nature of the sample.  Although Schmidt-Kaler
(1982) does not list the {\it (B--V)$_{0}$} for K0IV objects, the
interpolated value between K0V and K0III stars is 0.90, close to what is
measured.  The remaining 6 objects have even redder colours, {\it
(B--V)}~=~1.2~$\pm$~0.2 with an average \Mv~=~0.1~$\pm$~1 mag,
suggesting that these may be K0III stars. 

The simplest explanation for the large range in absolute magnitudes
then appears that the sample of K0V stars in the MSS survey is
contaminated by K0IV objects (about 20\% of the entire
sample), and perhaps suffers from contamination from 
intrinsically even brighter objects.

\begin{figure}
\mbox{\epsfxsize=0.46\textwidth\epsfbox[32 171 512 425]
{
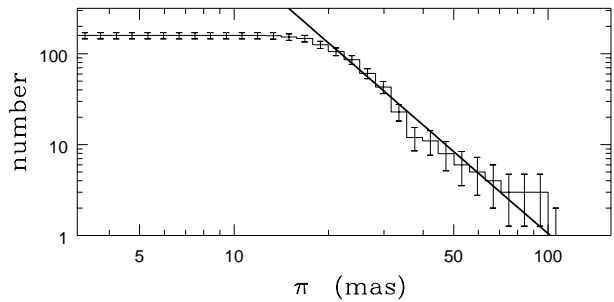}}
\caption{ 
Cumulative parallax distribution of the K0V sample. The solid line
indicates a fit through the data points with $\pi$~$>$~20 mas. The bins
are 0.05 wide in log($\pi$) units, the errorbars represent the
statistical error ($\sqrt{N}$).
\label{kovf6}
}
\end{figure}

\section{The K0V subsample and the Malmquist bias}

In the following, we will continue with the K0V sample, i.e. the 159
objects with inferred intrinsic magnitudes fainter than 4.5.  Our 
interest is whether the `completeness effect' or Malmquist
bias affects the determination of  the intrinsic magnitude of the
sample under consideration.
Fig.~\ref{kovf3} shows the same figure as the lower panel of
Fig.~\ref{kovf1}, but now only for the K0V sample.  As before, there
is a clear trend visible.  The smaller the parallax, the 
brighter the mean is.  We have binned the data in steps of $\pi$,
and calculated the mean and its scatter.  In the
interval $\pi$~=~10-12.5 mas (which are still 6-10$\sigma$
detections), the mean  is 0.7 mag brighter than in
the interval 60-100 mas.

The change in mean absolute magnitude is easily understood.  This can
be learned from the volume completeness of the sample.
Fig~\ref{kovf6} shows the cumulative parallax distribution of the
stars.  The distribution flattens below $\pi$~$<$~20 mas indicating
that the sample is complete to $\approx$~20 mas.  The solid line
represents a weighted least-squares fit to the data between 20 and 100
mas, with   slope   $-3\pm$0.15, implying a uniform space
distribution, consistent with the small volume probed to 50 pc.  The
Malmquist bias only occurs for volume-incomplete samples, and indeed,
for $\pi$~$>$~20 mas, the mean magnitude in the bins does not change
in Fig.~\ref{kovf3}, but it does for the lower parallax values.

What is the effect {\it in this particular case} on the derived absolute
magnitudes of K0V stars if the Malmquist bias would not have been taken
into account? The unweighted mean of the 159 objects is 5.59~$\pm$~0.42
mag, while the unweighted mean for the unaffected sample (85 stars with
$\pi$~$>$~20 mas) is \Mv~=~5.69~$\pm$~0.40 mag.  The scatter of
$\approx$~0.40 reflects the intrinsic scatter rather than errors arising
from the measurement uncertainties, and may for example be due to
variations in metallicity, rotation period or unseen binaries.  The
difference between the two values is more in agreement with the
prediction that the Malmquist bias is of order 0.2 mag -- this is
dependent on the distinction between the K0V and the `K0IV'
samples, because a fainter cut-off value results in a slightly smaller
scatter around the mean.

However, there is one significant difference with e.g.  the situation of
red-shifts as distance determinations: in the parallax case the relative
observational error \spi , is much larger than in the case of the
red-shifts, so a straightforward averaging of the derived absolute
magnitudes should be replaced by a weighting scheme.  This will decrease
the effect of the Malmquist bias somewhat: The objects that are more
prone to the selection effects are further away, and have larger
relative errors on the parallax, they will therefore have less weight. 
Since the error is asymmetric in magnitudes, we now have to calculate
the weighted mean in `reduced parallax' (10$^{0.2M_{V}}$).  The
weighting of all 159 objects now results in a mean \Mv~=~5.65, while
the 85 objects with $\pi$~$>$~20 mas have a weighted mean of 5.69 mag. 
So, for this sample, consisting of both high quality parallax
measurements and spectral types, not taking into account the Malmquist
bias would result in an unweighted mean too bright consistent with the
expected value of the Malmquist bias, and a weighted mean
absolute magnitude that is too bright by 0.04 mag. 

The main result concerning the `true' intrinsic magnitude of K0V stars
is that the sample which is not affected by contamination by K0IV stars
yields a value 0.2 mag brighter than has been listed in the literature
so far (SK82).  The existing calibrations apparently need a revision and
this work illustrates the power of Hipparcos trigonometric parallaxes. 
An additional result is that the stars that we tentatively identify as
K0IV objects seem to be 0.4 mag.  fainter than expected.  However, we do
not put much weight to this result, as these by implication would be
mis-classified K0V stars, and thus likely to be those K0IV objects that are
on the fainter side of the distribution in the first place.  A detailed
study of this effect is beyond the scope of the present paper.

\section{Final remarks}

We have studied a sample of 200 K0V stars,  taken from
the best collection of spectral types available, the Michigan Spectral
Survey, which have excellent trigonometric parallaxes from the
Hipparcos mission.  In our high quality data, the Malmquist bias occurs
already when the measured parallax is 20 mas, the point where
volume-incompleteness sets in.  The presence of the Malmquist bias is
readily seen, when the Spaenhauer diagram is used as a diagnostic tool. 
Only because K0V stars are intrinsically faint, the Malmquist bias
starts to play a role this quickly.  For intrinsically brighter objects,
or samples with  fainter limiting {\it V} magnitudes, the
completeness limit will be pushed to lower observed parallaxes.

So far, we have not discussed the Lutz-Kelker (1973) effect -- as
shown in OGS98, this effect becomes important when the relative error
on the parallax is still comparitively low (10-20\%).  Since the
absolute errors on the parallaxes are almost all the same
($\approx$~1.4 mas in our sample), this corresponds to an observed
$\pi$ of 7-14 mas.  In the case of the faint K0V stars, the Malmquist
effect dominates at these parallaxes, so objects that otherwise would
have had too faint derived absolute magnitudes are excluded from the
sample, perhaps lowering the effect of the Malmquist bias.  To assess
such effects, one has to examine intrinsically brighter objects, as
done in OGS98, or by Kaltcheva \& Knude (1998) who investigated B
stars.  The latter authors showed that for well determined parallaxes
(\spi~$<$~10\%), the absolute magnitudes of B stars derived from the
parallax agree with the absolute magnitudes derived from H$\beta$
photometric distances, but for less well determined parallaxes (\spi \
between 10\% and 20\%), the individual objects are too faint, in
agreement with the prediction for Lutz-Kelker bias.

As a final comment, it is expected that each sample of stars will prove to
be sensitive to the Malmquist and Lutz-Kelker biases in its own unique
way, and only careful examination of the data can make their, sometimes
hidden, effects visible.  An additional result of this work is that
around 41 out of 200 K0V stars may be misclassified K0IV stars. 

\paragraph{Acknowledgements} 

RDO is funded by PPARC. This paper is based on data from the ESA Hipparcos
 satellite.

{}

\end{document}